\documentclass[11pt,a4paper]{article}
\usepackage{authblk}
\usepackage[hyperref]{emnlp2020}
\usepackage{times}
\usepackage{latexsym}
\usepackage{graphicx}
\usepackage{xspace}
\usepackage{subcaption}
\usepackage{makecell}
\usepackage{algorithm}
\usepackage{algorithmic}
\usepackage{array}
\usepackage{setspace}
\newcolumntype{C}[1]{>{\centering}p{#1}}
\usepackage{float}
\usepackage{url}
\usepackage{balance}
\usepackage{amsmath}
\usepackage{enumitem}
\usepackage{multirow}
\usepackage{color,xcolor}
\usepackage{breakcites}

\usepackage{microtype}

\aclfinalcopy 

\newcommand{\our}{\textsc{EvidenceMiner}\xspace}

\title{Automatic Textual Evidence Mining in COVID-19 Literature}

\author{
	\textbf{Xuan Wang}$^{1}$,
	\textbf{Weili Liu}$^{1}$,
	\textbf{Aabhas Chauhan}$^{1}$,
	\textbf{Yingjun Guan}$^{2}$,
	\textbf{Jiawei Han}$^{1}$ \\
	$^{1}$Department of Computer Science, University of Illinois at Urbana-Champaign \\
	$^{2}$School of Information Sciences, University of Illinois at Urbana-Champaign \\
	$^{1,2}$\texttt{\{xwang174,weilil2,aabhasc2,yingjun2,hanj\}@illinois.edu}
}

\date{}

\begin{document}
	\maketitle
	\begin{abstract}
		We created this \our system for automatic textual evidence mining in COVID-19 literature. \our is a web-based system that lets users query a natural language statement and automatically retrieves textual evidence from a background corpora for life sciences. 
		It is constructed in a completely automated way without any human effort for training data annotation. \our is supported by novel data-driven methods for distantly supervised named entity recognition and open information extraction. 
		The named entities and meta-patterns are pre-computed and indexed offline to support fast online evidence retrieval. 
		The annotation results are also highlighted in the original document for better visualization.
		\our also includes analytic functionalities such as the most frequent entity and relation summarization.
	\end{abstract}

	\section{Introduction}
	Coronavirus disease 2019 (COVID-19) is an infectious disease caused by severe acute respiratory syndrome coronavirus 2 (SARS-CoV-2). The disease was first identified in 2019 in Wuhan, Central China, and has since spread globally, resulting in the 2019–2020 coronavirus pandemic.
	On March 16th, 2020, researchers and leaders from the Allen Institute for AI, Chan Zuckerberg Initiative (CZI), Georgetown University's Center for Security and Emerging Technology (CSET), Microsoft, and the National Library of Medicine (NLM) at the National Institutes of Health released the COVID-19 Open Research Dataset (CORD-19)\footnote{\url{https://www.kaggle.com/allen-institute-for-ai/CORD-19-research-challenge}} of scholarly literature about COVID-19, SARS-CoV-2, and the coronavirus group.
	
	Traditional search engines for life sciences (e.g., PubMed) are designed for document retrieval and do not allow direct retrieval of specific statements. Some of these statements may serve as textual evidence that is key to tasks such as hypothesis generation and new finding validation. 
	We created \our \cite{wang2020evidenceminer}, a web-based system for textual evidence discovery for life sciences. We apply the \our system to the CORD-19 corpus to facilitate textual evidence mining for the COVID-19 studies\footnote{\url{https://xuanwang91.github.io/2020-04-15-cord19-evidenceminer/}}. Given a query as a natural language statement, \our automatically retrieves sentence-level textual evidence from the CORD-19 corpus.
	In the following sections, we introduce the details of the \our system. We also show some textual evidence retrieval results with \our on the CORD-19 corpus.
	
	\begin{figure}[t]
		\includegraphics[width=0.45\textwidth]{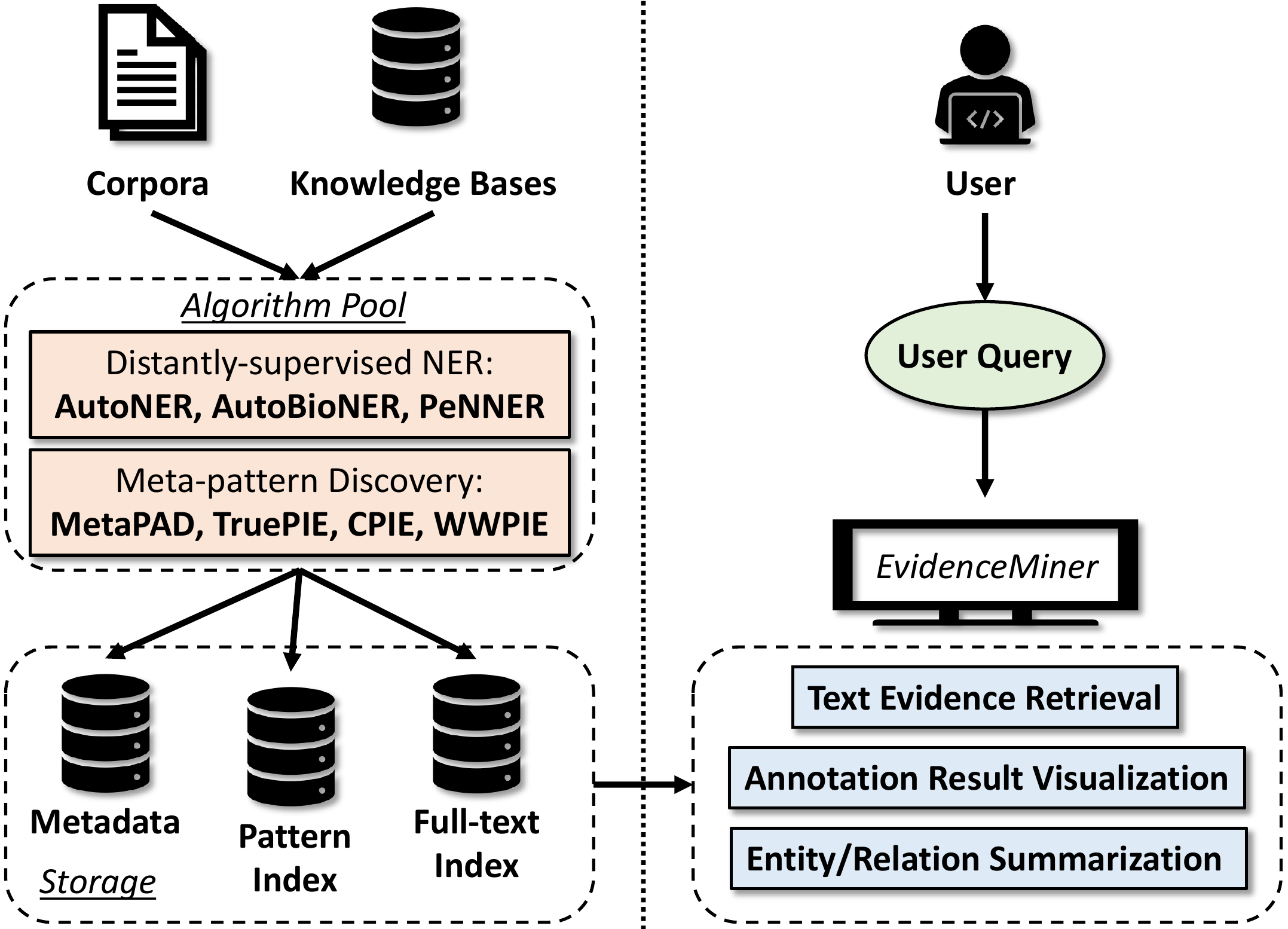}
		\caption{System architecture of \our.}
		\label{fig:architecture}
	\end{figure}

	\section{\our System}
	\our is a web-based system for textual evidence discovery for life sciences (Figure \ref{fig:architecture}). Given a query as a natural language statement, \our automatically retrieves sentence-level textual evidence from a background corpora of biomedical literature. \our is constructed in a completely automated way without any human effort for training data annotation. It is supported by novel data-driven methods for distantly supervised named entity recognition and open information extraction. 
	
	\subsection{Pre-processing}
	\our relies on external knowledge bases to provide distant supervision for named entity recognition (NER) \cite{shang2018learning, wang2018penner, wang2019distantly}. For this COVID-19 study, the NER results are obtained from the CORD-NER system \cite{wang2020comprehensive}. Based on the entity annotation results, it automatically extracts informative meta-patterns (textual patterns containing entity types, e.g., CHEMICAL inhibit DISEASE) from sentences in the background corpora. \cite{jiang2017metapad, wang2018open, li2018truepie, li2018pattern}. Sentences with meta-patterns that better match the query statement is more likely to be textual evidence. 
	
	\subsection{Corpus Indexing}
	After we get all entities and meta-patterns extracted, we use them to guide the textual evidence discovery. We create three offline indexes of the input corpus based on the words, entities and synonym meta-pattern groups from the previous steps. Indexing helps boost the online processing of textual evidence discovery given a user-specified query.
	
	\smallskip \noindent \textbf{Word indexing.}
	We normalize the words in the corpus to lowercase and split the corpus into single sentences for indexing. We take each word in the vocabulary of the corpus as the key and generate an identifier list as the value including the sentences containing the key word. 
	
	\smallskip \noindent \textbf{Entity indexing.}
	Similar to word indexing, we take each entity recognized by PubTator as the key and generate an identifier list as the value including the sentences containing the key entity. 
	
	\smallskip \noindent \textbf{Pattern indexing.}
	After the fine-grained meta-pattern matching, we take each meta-pattern together with its extracted entity set as two levels of keys and generate the identifier list as the value including the sentences matched by the meta-patterns with their corresponding entity set. We also maintain two dictionaries, one mapping each meta-pattern to its synonym meta-pattern group and the other mapping each synonym meta-pattern group to its list of meta-patterns.
	
	\subsection{Textual Evidence Retrieval.}
	Taken an user-input query and the indexed corpus, we retrieve all the candidate evidence sentences and rank them by a confidence score of it being textual evidence for the query. The confidence score is a weighted combination of three scores: a word score, an entity score and a pattern score. 
	Our evidence ranking function calculates the following three ranking scores:
	\begin{enumerate}
		\item Word score: candidate evidence sentences covering more query-related words will be ranked
		higher.
		\item Entity score: candidate evidence sentences covering more query-related entities will be ranked higher.
		\item Pattern score: candidate evidence sentences covering more query-matched meta-patterns will be ranked higher.
	\end{enumerate}
	
	
	\smallskip \noindent \textbf{Word Score.} We use the BM25 \cite{robertson2009probabilistic} score as the word score to measure the relatedness between the query and the candidate evidence sentence. BM25 is a commonly used ranking score for information retrieval. Given a query $Q$ containing the words $(w_1, w_2, ..., w_n)$, the BM25 score of a candidate evidence sentence $S$ is
	
	{\small
		\begin{align*}
		& S_w (Q, S) \\
		= & \sum_{i=1}^n IDF(w_i)\cdot\frac{f(w_i, S)\cdot(k+1)}{f(w_i, S) + k\cdot(1-b+b\cdot\frac{|S|}{avgsl})}, 
		\end{align*}
	}%
	
	where $f(w_i, S)$ is the term frequency of $w_i$ in the sentence $S$, $|S|$ is the length of the sentence $S$, $avgsl$ is the average length of all the sentences and k and b are two free parameters chosen by the user. $IDF(w_i)$ is the inverse document frequency of $w_i$, which is computed as
	
	{\small
		\begin{align*}
		IDF(w_i) = log\frac{N-n(w_i)+0.5}{n(w_i)+0.5},
		\end{align*}
	}%
	
	where $N$ is the total number of sentences and $n(w_i)$ is the number of sentences containing $w_i$. A candidate evidence sentence that is more related to the query will have a higher word score.
	
	\smallskip \noindent \textbf{Entity Score.}
	We also use the BM25 score as the entity score to measure the relatedness of the query and the candidate evidence sentence. Given a query $Q$ containing the entities $(e_1, e_2, ..., e_m)$, the BM25 score of a candidate evidence sentence $S$ is
	
	{\small
		\begin{align*}
		& S_E (Q, S) \\
		= & \sum_{i=1}^m IDF(e_i)\cdot\frac{f(e_i, S)\cdot(k+1)}{f(e_i, S) + k\cdot(1-b+b\cdot\frac{|S|}{avgsl})}, 
		\end{align*}
	}%
	
	where $f(e_i, S)$ is the term frequency of $e_i$ in the sentence $S$, $|S|$ is the length of the sentence $S$, $avgsl$ is the average length of all the sentences and k and b are two free parameters chosen by the user. $IDF(e_i)$ is the inverse document frequency of $e_i$, which is computed as
	
	{\small
		\begin{align*}
		IDF(e_i) = log\frac{N-n(e_i)+0.5}{n(e_i)+0.5},
		\end{align*}
	}%
	
	where $N$ is the total number of sentences and $n(e_i)$ is the number of sentences containing $e_i$. A candidate evidence sentence more related to the query will have a higher entity score.
	
	\smallskip \noindent \textbf{Pattern Score.} 
	We measure how many synonym patterns to the query pattern can be matched on each candidate evidence sentence. Given an input query (e.g., \textit{(resveratrol, inhibit, pancreatic cancer)}), we first try to convert it into a query meta-pattern (e.g., ``\$CHEMICAL inhibit \$DISEASE"). If the query meta-pattern can be found in our pattern index, we directly retrieve all the synonym meta-patterns to the query meta-pattern. Then we measure how many meta-patterns among the synonym meta-patterns can be matched for each candidate evidence sentence on the query entities. Given a query $Q$ containing the entities $(e_1, e_2, ..., e_n)$, the pattern score of a candidate evidence sentence $S$ is
	
	{\small
		\begin{align*}
		S_P (Q, S) = \sum_{i=1}^k Match(MP_i(Q), S),
		\end{align*}
	}%
	
	where $k$ is the number of meta-patterns in the synonym meta-pattern group of the query on $Q$, $MP_i(S)$ is each meta-pattern in the synonym meta-pattern group of the query on $Q$, and $Match(MP_i(Q), S)$ is an indicator function that measures if the sentence $S$ can be matched with $MP_i(Q)$ on the query entities.
	A candidate evidence sentence is more likely to be confident evidence if it can be matched to more synonym meta-patterns to the query meta-pattern.
	
	\smallskip \noindent \textbf{Textual Evidence Score.}
	The final score of the candidate evidence sentence is a weighted average of the three scores,
	
	{\small
		\begin{align*}
		S(Q, S) = \sigma \cdot S_w + \theta \cdot S_E + \eta \cdot S_P, 
		\end{align*}
	}%
	
	where $(\sigma, \theta, \eta)$ is the weight vector indicating the importance of each aspect of the information (i.e., word, entity and pattern), which can be adjusted by the user. The default weight vector we use is equal weight for word, entity and pattern in our experiments.

	\section{Results on COVID-19}
	
	\subsection{Textual Evidence Retrieval}
	To demonstrate the effectiveness of \our in textual evidence retrieval, we compare its performance with the traditional BM25 \cite{robertson2009probabilistic} and a recent sentence-level search engine, LitSense \cite{allot2019litsense}. The background corpus is the same PubMed subset for all the compared methods. We first ask domain experts to generate 50 query statements based on the relationships between three biomedical entity types (gene, chemical, and disease) in the Comparative Toxicogenomics Database\footnote{\url{http://ctdbase.org}}. Then we ask domain experts to manually label the top-10 retrieved evidence sentences by each method with three grades indicating the confidence of the evidence.
	We use the average normalized Discounted Cumulative Gain (nDCG) score to evaluate the textual evidence retrieval performance.
	In Table \ref{tab:ranking}, we observe that \our always achieves the best performance compared with other methods. 
	It demonstrates the effectiveness of using entities and meta-patterns to guide textual evidence discovery in biomedical literature. 
	
	\begin{table}[t]
		\centering
		\small
		\begin{tabular}{c|c|c|c}
			\hline
			\textbf{Method /\ nDCG} & \textbf{@1} & \textbf{@5} & \textbf{@10} \\
			\hline
			BM25 & 0.714 & 0.720 & 0.746 \\
			LitSense & 0.599 & 0.624 & 0.658 \\
			\our & \textbf{0.855} & \textbf{0.861} & \textbf{0.889}\\
			\hline
		\end{tabular}
		\caption{Performance comparison of the textual evidence retrieval systems with nDCG@1,5,10.}
		\label{tab:ranking}
	\end{table}

	\subsection{Case Study}
\begin{figure*}[t]
		\centering
		\includegraphics[width=\textwidth]{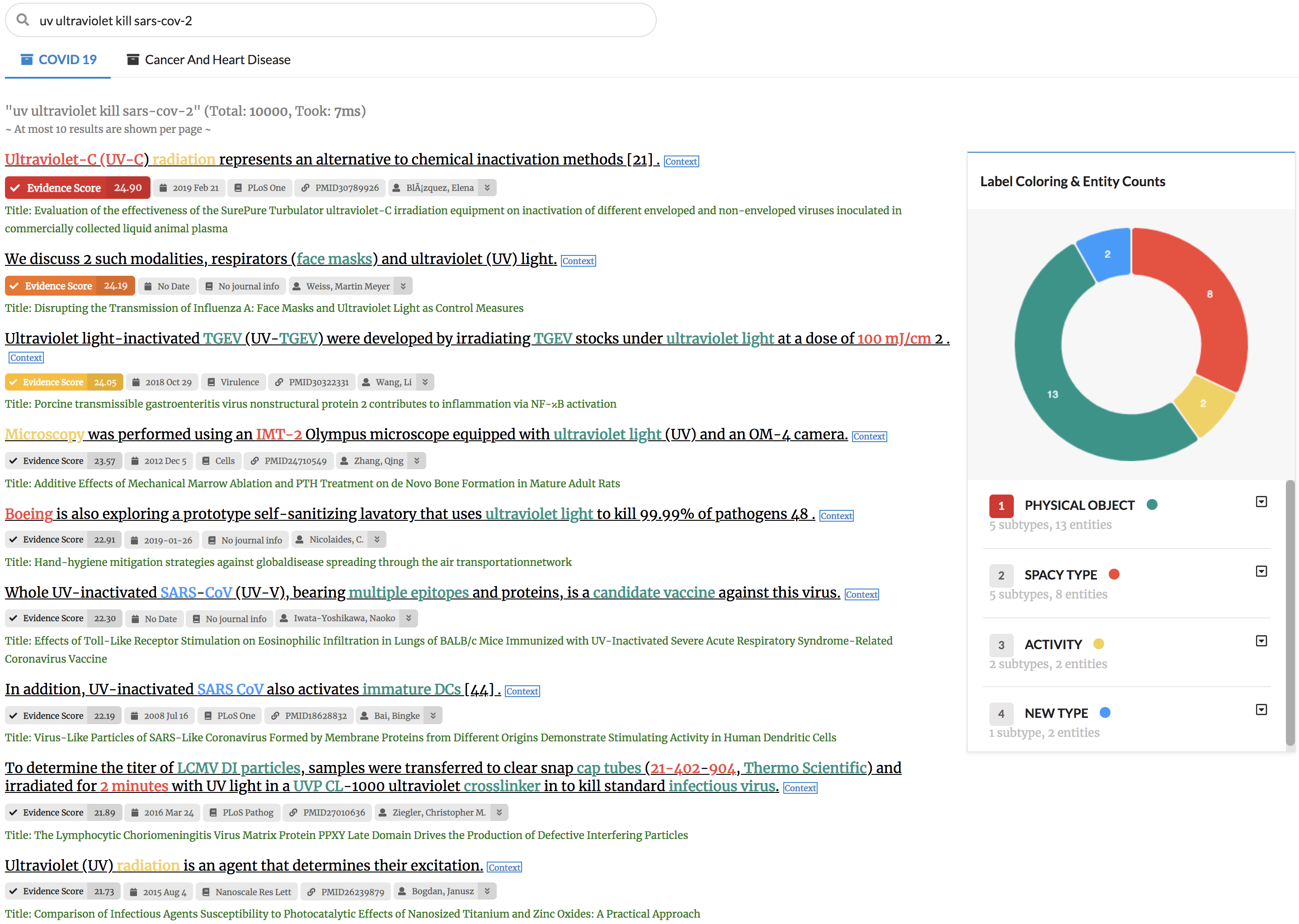}
		\caption{Case study: (Ultraviolet, UV, kills, SARS-COV-2)}
		\label{fig:case-a}
	\end{figure*}

	Here are some case studies to demonstrate that \our can help scientific discoveries on COVID-19. In Figure \ref{fig:case-a}, scientists want to find some evidence for using ultraviolet (UV) to kill the SARS-COV-2 virus. In the top-retrieved results, we see many supporting sentences such as the top one “Ultraviolet-C (UV-C) radiation represents an alternative to chemical inactivation methods”. More interestingly, we found the fifth sentence “Whole UV-inactivated SARS-CoV (UV-V), bearing multiple epitopes and proteins, is a candidate vaccine against this virus” indicating that UV-inactivation also has the potential for vaccine development against the virus. Scientists are very interested in this result that inspired them to conduct UV-related COVID-19 vaccine development.
	
\begin{figure*}[t]
		\centering
		\includegraphics[width=\textwidth]{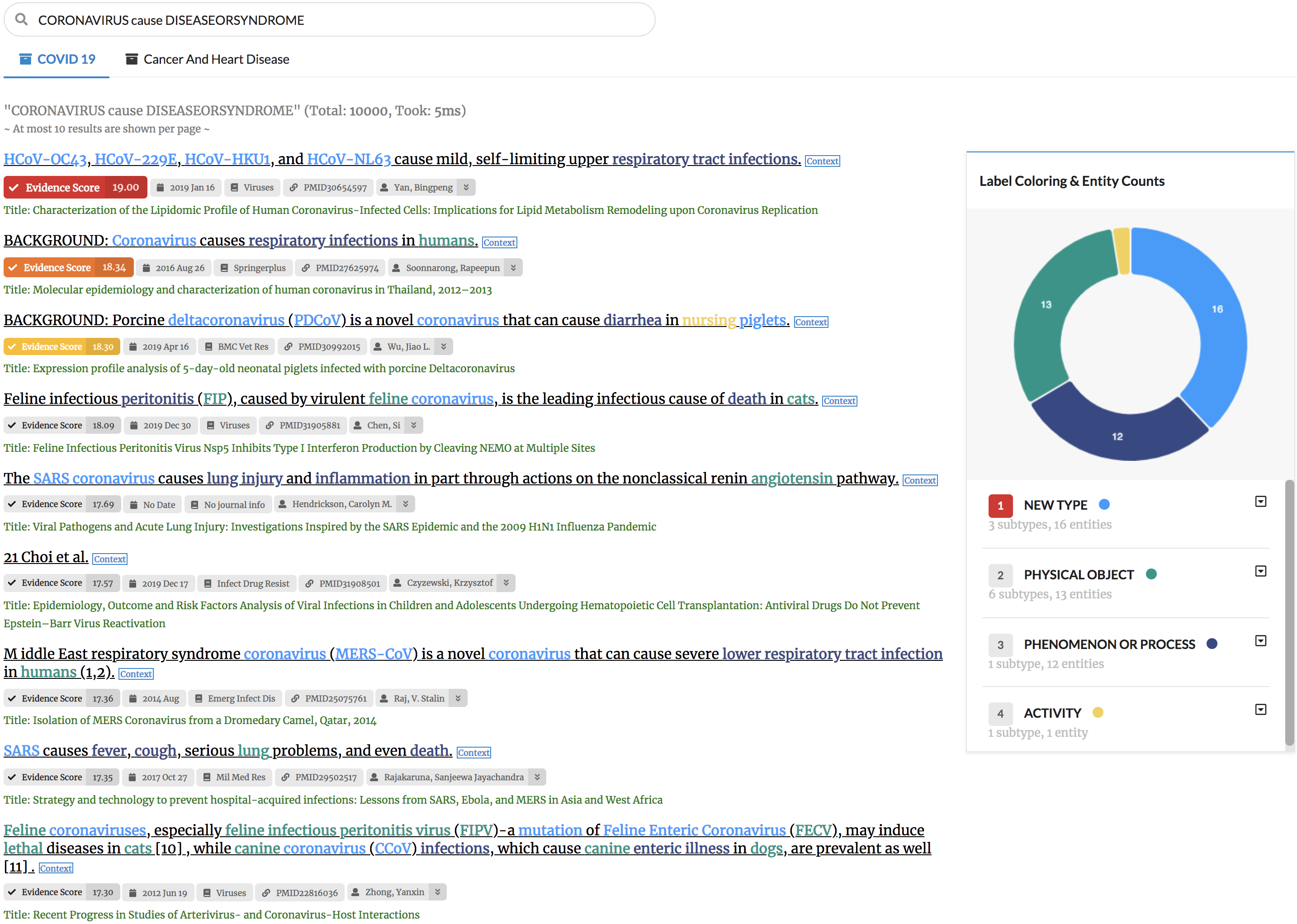}
		\caption{Case study: (CORONAVIRUS cause DISEASEORSYNDROME)}
		\label{fig:case-b}
	\end{figure*}

	Moreover, \our allows more flexible queries, such as the relational patterns, if the users are not sure which specific entity to search. In Figure \ref{fig:case-b}, scientists want to find some evidence related to “CORONAVIRUS cause DISEASEORSYNDROME”. In the top-retrieved results, we see many highly-related evidence sentences, such as “HCoV-OC43, HCoV-229E, HCoV-HKU1, and HCoV-NL63 cause mild, self-limiting upper respiratory tract infections”. This function is supported by our meta-pattern discovery methods and has not been incorporated by any existing systems.
	
\begin{figure*}[t]
		\centering
		\includegraphics[width=\textwidth]{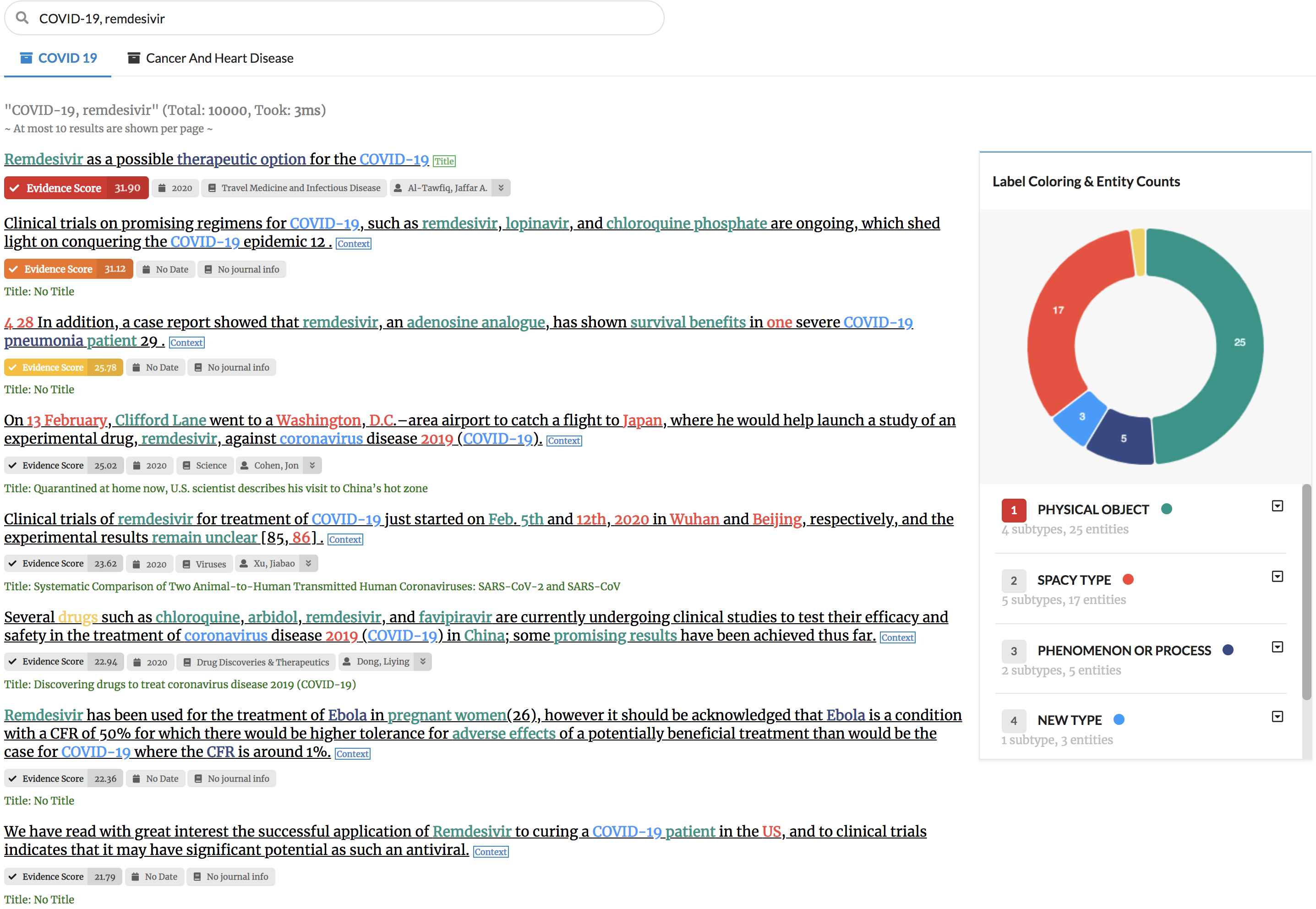}
		\caption{Case study: (COVID-19, remdesivir)}
		\label{fig:case-c}
	\end{figure*}

\begin{figure*}[t]
		\centering
		\includegraphics[width=\textwidth]{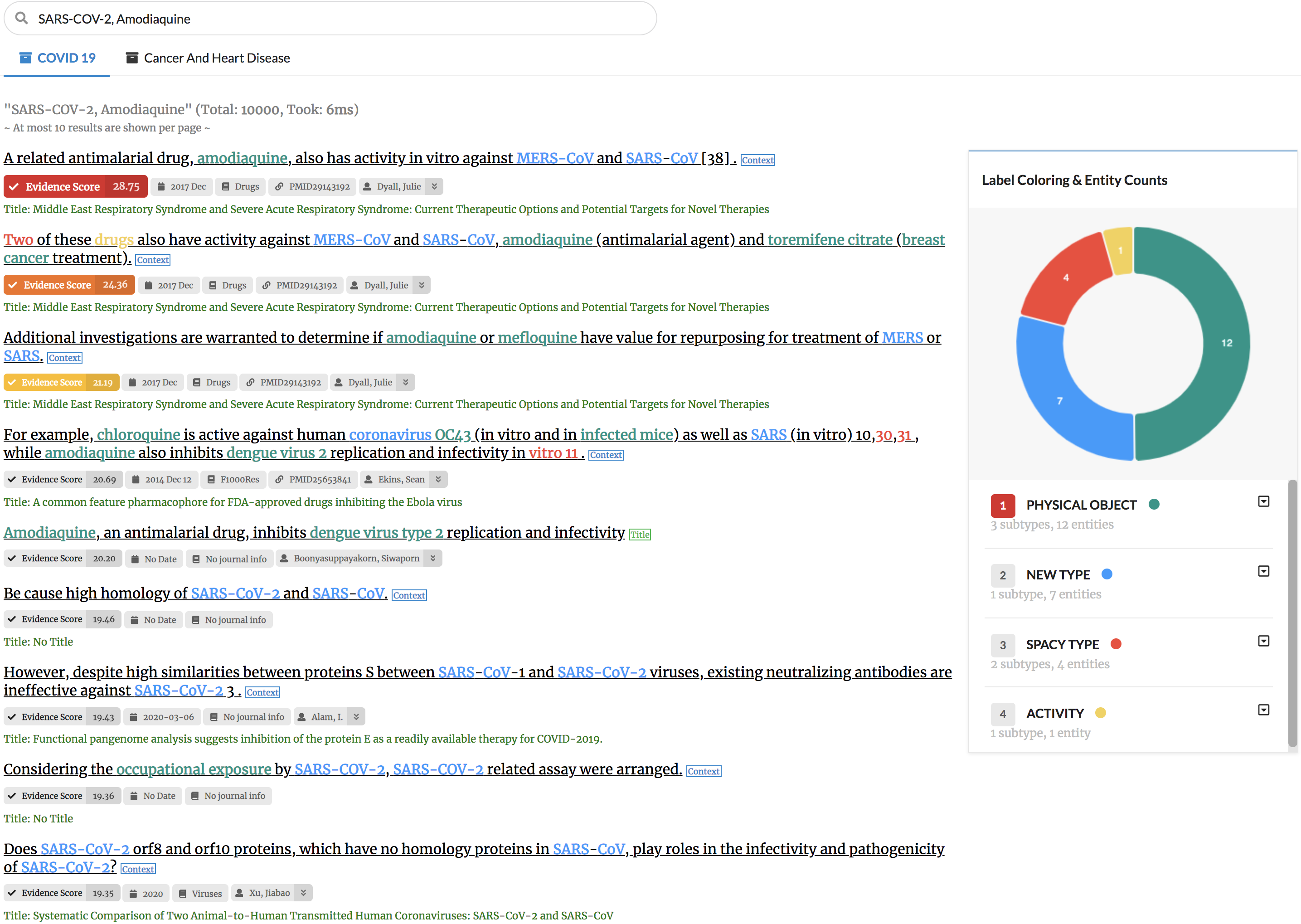}
		\caption{Case study: (COVID-19, amodiaquine)}
		\label{fig:case-d}
	\end{figure*}

	We show some more examples. In Figure \ref{fig:case-c}, doctors want to study if remdesivir is a potential drug treatment for COVID-19. Remdesivir is currently a very actively studied drug that has the potential to be repurposed for COVID-19 treatment. Similarly, in the top-retrieved results shown below, we can see many sentences regarding the clinical trials for remdesivir against COVID-19. An additional example is shown for amodiaquine as a potential drug for COVID-19 in Figure \ref{fig:case-d}.
	
\begin{figure*}[t]
		\centering
		\includegraphics[width=\textwidth]{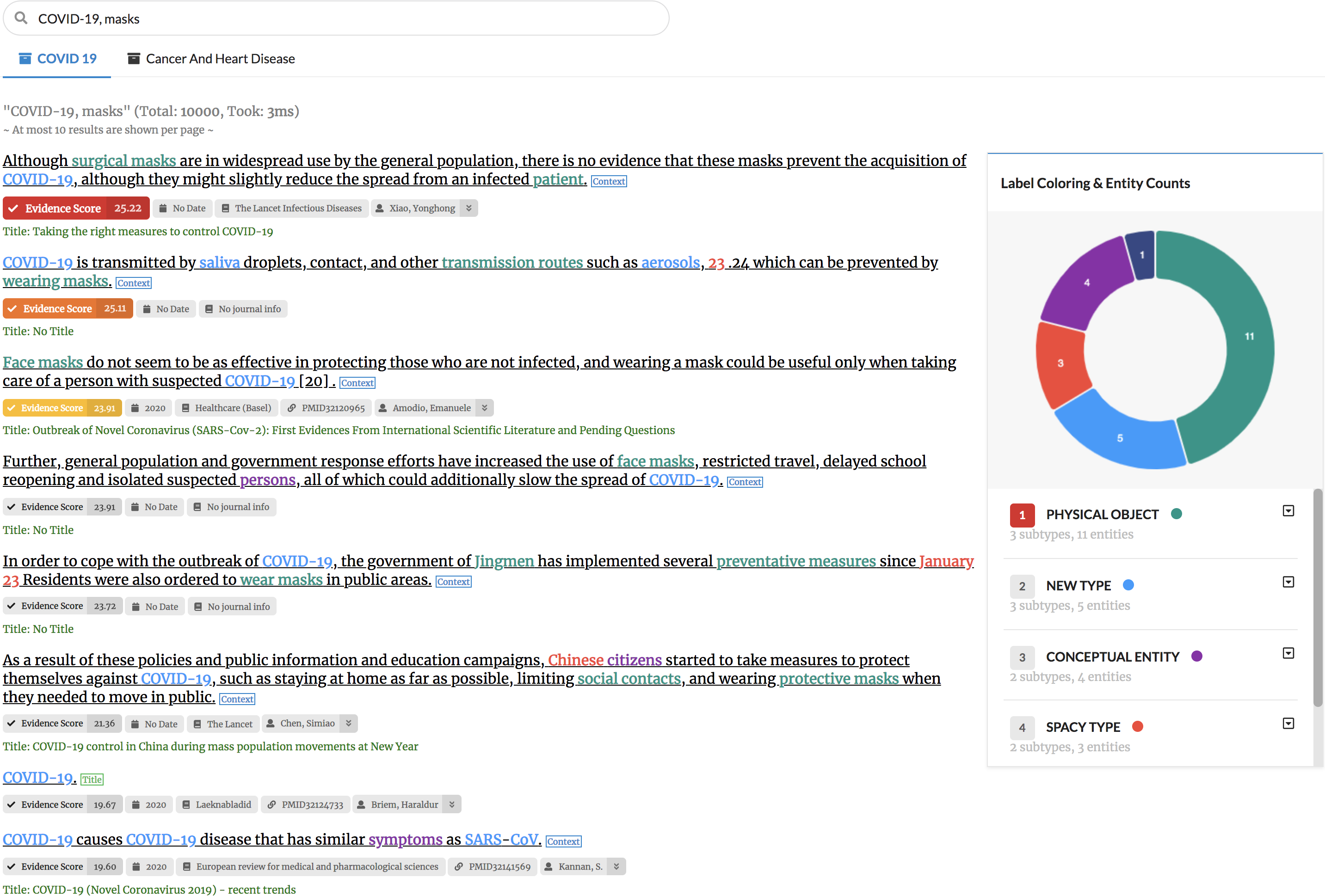}
		\caption{Case study: (COVID-19, masks)}
		\label{fig:case-e}
	\end{figure*}

	Last, we show that \our is also useful for evidence finding for controversial topics. In Figure \ref{fig:case-e}, people are interested to see if wearing masks can help prevent the COVID-19 spreading. In the top-retrieved results, we see many related statements, among them are clearly two opposite opinions. For example, some statements support the use of masks to prevent the virus, such as “COVID-19 is transmitted by saliva droplets, …, which can be prevented by wearing masks”. While other statements are against the effectiveness of wearing masks, such as “Although surgical masks are in widespread use …, there is no evidence that wearing these masks can prevent the acquisition of COVID-19 …”. An interesting future work is to classify the opinions by their semantic polarity and even automatically generate summarizations of the evidence retrieval results.

	\section{Conclusion}
	\our on COVID-19 be constantly updated based on the incremental updates of the CORD-19 corpus and the improvement of our system. We hope this system can help the text mining community build downstream applications for the COVID-19 related tasks. We also hope this system can bring insights for the COVID-19 studies on making scientific discoveries.

	\section*{Acknowledgment}
	Research was sponsored in part by US DARPA KAIROS Program No. FA8750-19-2-1004 and SocialSim Program No.  W911NF-17-C-0099, National Science Foundation IIS 16-18481, IIS 17-04532, and IIS-17-41317, and DTRA HDTRA11810026. Any opinions, findings, and conclusions or recommendations expressed herein are those of the authors and should not be interpreted as necessarily representing the views, either expressed or implied, of DARPA or the U.S. Government. The U.S. Government is authorized to reproduce and distribute reprints for government purposes notwithstanding any copyright annotation hereon. The views and conclusions contained in this paper are those of the authors and should not be interpreted as representing any funding agencies.

	\bibliography{anthology,emnlp2020}
	\bibliographystyle{acl_natbib}
	
\end{document}